\documentclass[preprint,aps,draft]{revtex4}

\usepackage{bm}

\begin{document}


\title{Finite number of Kaluza-Klein modes, all with zero masses}

\author{Recai Erdem}
\email{recaierdem@iyte.edu.tr}
\affiliation{Department of Physics,
{\.{I}}zmir Institute of Technology \\ 
G{\"{u}}lbah{\c{c}}e K{\"{o}}y{\"{u}}, Urla, {\.{I}}zmir 35430, 
Turkey} 

\date{\today}

\begin{abstract}
Kaluza-Klein modes of fermions in a 5-dimensional toy model are 
considered. The number of Kaluza-Klein modes that survive after 
integration over extra dimensions is finite in this space. Moreover the 
extra dimensional piece of the kinetic part of the Lagrangian in this 
space induces no mass for the higher Kaluza-Klein modes on contrary to the 
standard lore.
 \end{abstract}

\maketitle

The use of extra dimension(s) is a popular tool in high energy physics 
\cite{Kaluza,Klein, Duff, ED1,Antoniadis0,ED2,RS1,RS2,Uehara,Flacke} 
because it gives a more tidy picture of nature, that 
ranges from geometrization of  all forces of nature 
in the spirit of general relativity to a better understanding 
of the cosmological constant problem, hierarchy problem, fermion 
generations, 
Yukawa couplings and 
flavor etc.. The world we live in is apparently 4-dimensional. Hence extra 
dimensions (if exist) must be hidden at present (relatively low) 
energies. The standard way to ensure this is to take the extra 
dimension(s) be compact and tiny (e.g. a tiny circle). Then, by Fourier 
theorem, a field in the whole space can be expanded in a tower of 
particles that are identical except their masses and their profile in the 
extra dimension(s). Such a tower of a particle (or field) is called a 
Kaluza-Klein (KK) tower of that particle (or field), and it is an infinite 
series except in some exceptional cases 
that need complicated boundary conditions to be satisfied    
\cite{Grard-Nuyts1,Grard-Nuyts2,Grard-Nuyts3,Grard-Nuyts4}. 
Depending on the boundary conditions 
the KK tower may contain a zero mode (i.e. a mode 
that does not depend on the extra dimension(s)) or not. A zero mode does 
not acquire a mass from the extra dimensional piece of the kinetic part of 
the Lagrangian while all other modes gain masses of order of $\frac{1}{L}$ 
where $L$ is the size of the extra dimension. Phenomenological 
considerations require the masses of the higher KK modes be at 
least in $TeV$ scale, and usually in the order of Planck mass for 
standard model particles \cite{Antoniadis}. 
So KK modes except the zero mode can not be 
identified with the usual particles. Therefore a scheme where the number 
of KK modes is finite and all gain zero masses from the kinetic part 
of the Lagrangian would be highly desirable. I this study I present a toy 
model for fermions where the number of observed KK modes is finite at 
current energies, and all modes are massless as long as the kinetic part 
of the Lagrangian is considered.  

In the vein of a framework proposed for cosmological constant 
problem \cite{Erdem1, Erdem2,Erdem3,Erdem4} I consider the following 
5-dimensional 
metric 
\begin{eqnarray}
ds^2&=&
g_{BC}\,dx^B dx^C
\,=\,\cos{kz}\,\left(
g_{\bar{B}\bar{C}}\,dx^{\bar{B}} dx^{\bar{C}}\right) \nonumber \\
&=&\cos{k\,z}
\left[\,g_{\bar{\mu}\bar{\nu}}\left(x\right)\,dx^{\bar{\mu}} 
dx^{\bar{\nu}}
-\,dz^2\,\right]
\label{a1} \\
&&B,C,\bar{B},\bar{C}\,=\,0,1,2,3,4~~~,~~~~~\bar{\mu},\bar{\nu}\,=\,0,1,2,3 
\nonumber 
\end{eqnarray}
where the symbol $x$ with no indices stands for the 4-dimensional 
coordinates $x^{\bar{\mu}}$. I take the extra dimension be compact and 
its 
size be $L$, and $k\,=\,\frac{2\pi}{L}$. Although this metric has 
singularity at $k\,z=\frac{\pi}{2}$ this singularity does not survive 
after integration over the extra dimension $z$ (i.e. at the scales larger 
than the size of the extra dimension). Moreover the location of the 
singularity at the sharp value, $k\,z=\frac{\pi}{2}$ suggests that this 
singularity may be removed by the metric fluctuations in quantum gravity 
\cite{Guendelman}. So given the toy model nature of this study I will not 
dwell on this technical point further for the sake of a relatively simple 
framework to study. 
  
I take 
$g_{\bar{B}\bar{C}}\,=\,\eta_{\bar{B}\bar{C}}\,=\,$diag$(1,-1,-1,-1,-1)$ 
(i.e. $g_{\bar{\mu}\bar{\nu}}\,=\,\eta_{\bar{\mu}\bar{\nu}}$)
to have a simple model where one can focus on the essential points of the 
model. The action for (free) fermionic fields for this space is 
\begin{eqnarray}
S_f &=& \int \left(\cos{kz}\right)^{\frac{5}{2}}{\cal L}_f\,d^4x\,dz
\nonumber \\
&=&
\int \left(\cos{kz}\right)^2\,i\bar{\chi}\gamma^{a}\left(\partial_a
\,+\,\frac{k}{8}\tan{kz}\,\left[\,\gamma_4\;,\;\gamma_{a}\right]\,\right)\chi 
\;d^4x\,dz\;+\,H.C.
\label{a2} \\
&&\left\{\gamma^a,
\gamma^b\right\}\,=\,2\eta^{ab}~~,~~\left(\eta^{ab}\right)=
\mbox{diag}\left(1,-1,-1,-1,-1\right) 
\nonumber
\end{eqnarray}
where $H.C.$ stands for Hermitian conjugate, and the second term 
is spin connection term (See Appendix A). The small Latin indices $a$, 
$b$, etc. refer to 
the tangent space while the capital Latin indices $A$,$B$, etc. 
refer to the space defined by (\ref{a1}). The tangent space 
in this case coincides with 
$g_{\bar{B}\bar{C}}\,dx^{\bar{B}}dx^{\bar{C}}$
=$\eta_{\bar{B}\bar{C}}\,dx^{\bar{B}}dx^{\bar{C}}$. So the indices with a 
bar on it also refer to the tangent space in this paper.
The action is required 
to be invariant under the 5-dimensional space-time reflections, namely,
\begin{eqnarray}
x^a\,\rightarrow\,-\,x^a ~~~~,~~~~~~ 
a\,=\,0,1,2,3,4
\label{a3}
\end{eqnarray}
where all coordinates are space-time reflected simultaneously.

$\chi$ may be Fourier decomposed in the extra dimension as
\begin{eqnarray}
\chi&=&\chi_{\cal A}\,+\,\chi_{\cal S} \label{a4aa} \\
\chi_{\cal A}\left(x,z\right)
&=&\sum_{n=-\infty}^{\infty}\,\chi^{\cal A}_n\left(x\right)\,
\,\sin{\left(\frac{1}{2}n\,kz\right)}\,=\,\sum_{|n|=1}^{\infty}
 \,\tilde{\chi}^{\cal A}_{|n|}\left(x\right)\,
\,\sin{\left(\frac{1}{2}|n|\,k z\right)} 
\label{a4a} \\
\chi_{\cal S}\left(x,z\right)
&=&\sum_{n=-\infty}^{\infty}\,\chi^{\cal S}_n\left(x\right)
\,\,\cos{\left(\frac{1}{2}n\,kz\right)}
\,=\,\chi_0\left(x\right)\,+\,\sum_{|n|=1}^{\infty}\,
\tilde{\chi}^{\cal S}_{|n|}(x)\,\,\cos{\left(\frac{1}{2}|n|\,k  z\right)} 
\label{a4b} \\
&&
\tilde{\chi}^{\cal A}_{|n|}\left(x\right)
\,=\,\chi^{\cal A}_n\left(x\right)\,-\,
\chi^{\cal A}_{-n}(x)~~,~~
\tilde{\chi}^{\cal S}_{|n|}\left(x\right)\,
=\,\chi^{\cal S}_n\left(x\right)\,+\,\chi^{\cal S}_{-n}(x) 
\nonumber
\end{eqnarray}
(where the absolute value signs in $|n|$ is used to emphasize the 
positiveness of $n$ in those terms, and half-fractional values in the sum 
correspond to anti-periodic boundary conditions). 
The form of 
$\chi_n^{{\cal A}\left({\cal S}\right)}$ is 
determined by the 
requirement of covariance under (the spinor representation of) 
$SO\left(3,1\right)$ 
and is given by 
\begin{eqnarray}
\chi_n^{{\cal A}
\left({\cal S}\right)}&=&\chi_{0\,n}^{{\cal A}
\left({\cal S}\right)}\,+\, 
\sum\,\Gamma^4\chi_{4\,n}^{{\cal A}\left({\cal S}\right)}
\label{a4c} \\
&&\left\{\Gamma^B,\Gamma^C\right\}\,=\,\frac{2}{\cos{kz}}\eta^{BC}~~~~
B,C=0,1,2,3,4 \nonumber
\end{eqnarray}
where $\Gamma^{B\left(C\right)}$'s are the gamma matrices of (\ref{a1}). 
However we take 
$\chi_n^{{\cal A}\left({\cal S}\right)}$ simply be 
$\chi_{0\,n}^{{\cal A}\left({\cal S}\right)}$ for the sake of simplicity 
and it does not 
essentially change the result as we shall mention when 
we discuss the masses of the KK modes.  Let us return to the main subject 
after this remark. 
I take $\chi_n^{{\cal A}\left({\cal S}\right)}$ to 
transform under (\ref{a3}) as
\begin{equation}
\chi_n\left(x\right)
~\rightarrow~\left(-1\right)^{\lambda_n}\,{\cal CPT}\,
\chi_n\left(-x\right)~~~~~,~~~~~\lambda_n=\frac{1}{2}
\left(-1\right)^{\frac{n}{2}}
\label{a6a}
\end{equation}
where the upper indices ${\cal A}$ and ${\cal S}$ are suppressed, and 
${\cal CPT}$ denotes the usual 4-dimensional CPT operator (acting on the 
spinor part of the field). Only the positions of fields (i.e. $x^a$'s) are 
multiplied by $-1$ while the orientation of the fields in the space-time 
remain essentially the same i.e. the spinor part of $\chi$ remains 
essentially the same. In this respect (\ref{a3}) is the analog of CPT 
transformation rather than PT transformation in 4-dimensions. The 
invariance of (\ref{a2}) under (\ref{a3}) requires 
$i\bar{\chi}\gamma^a\partial_a\chi$ in ${\cal L}_f$ be invariant under 
(\ref{a3}). 
$i\bar{\chi}\gamma^\mu\partial_\mu\chi$ is invariant under 4-dimensional 
CPT. These together imply that 
$i\bar{\chi}\gamma^\mu\partial_\mu\chi$ (i.e. 
$i\bar{\chi}\gamma^\mu\chi$) is even under 
the extra 
dimensional part of (\ref{a3}).
So the possible form of ${\cal L}_f$ (after requiring it be odd
under (\ref{a3})) is 
\begin{equation}
i\bar{\chi}_{\cal S}\gamma^{a}\partial_a\chi_{\cal 
S}~~~~,\mbox{and/or}~~~~ 
i\bar{\chi}_{\cal A}\gamma^{a}\partial_a\chi_{\cal A}  \label{a5}
\end{equation}
In other words (\ref{a3}) requires ${\cal L}_f$ be either of the terms in 
(\ref{a5}) or their linear combination.

Further the invariance of the action under an extra dimensional 
reflection similar to the one given in \cite{Erdem1,Erdem2}
\begin{equation}
k\,z\,\rightarrow\,\pi\,+\,k\,z \label{a6}
\end{equation}
is imposed. 
Under (\ref{a6}) 
the volume element in (\ref{a2}) transforms as 
\begin{equation}
\left(\cos{k\,z}\right)^{\frac{5}{2}}d^4x\,dz\,
\rightarrow\,\sqrt{-1}
\left(\cos{k\,z}\right)^{\frac{5}{2}}d^4x\,dz\, \label{a7}
\end{equation}
Then invariance of (\ref{a2}) under (\ref{a6}) requires
$i\bar{\chi}\gamma^{a}\partial_a\chi$ be even under the same 
transformation. I impose $\chi$ satisfy anti-periodic boundary 
conditions \cite{conformal} i.e. 
$\chi\left(z=0\right)=-\chi\left(z=L\right)$. 
This sets $n$ in (\ref{a4a},\ref{a4b}) 
be odd. Then, the invariance of action (including the quantum paths) under 
(\ref{a3}), (\ref{a6a}) and 
(\ref{a6}) requires the 4-dimensional part of $S_f$ be (see Appendix B)
\begin{eqnarray}
&&\sum_{r,s=0}^\infty \int \,d^4x
\,i\bar{\chi}_{\left(2|r|+1\right)}
\gamma^{\bar{\mu}}\partial_{\bar{\mu}}\chi_{\left(2|s|+1\right)}\nonumber \\
&&\times\,2\, \int\,dz\,\left(\cos{kz}\right)^2\,
\left[\cos{\frac{2|r|+1}{2}kz}\,
\cos{\frac{2|s|+1}{2}kz}
\,-\,\sin{\frac{2|r|+1}{2}kz}\,\sin{\frac{2|s|+1}{2}kz}\,
\right]\,
+\,H.C.\nonumber \\
&=&\sum_{r,s=0}^\infty \int \,d^4x
\;i\bar{\chi}_{\left(2|r|+1\right)}\gamma^{\bar{\mu}}\partial_{\bar{\mu}}
\chi_{\left(2|s|+1\right)}\,\int_0^{L}\,dz\,
\left(\cos{2kz}+1\right)\,
\cos{\left(|r|+|s|+1\right)kz}\,\,+\,H.C. \nonumber \\
&=&\frac{1}{2}\sum_{r,s=0}^\infty \,\int \,d^4x
\,i\bar{\chi}_{\left(2|r|+1\right)}\gamma^{\bar{\mu}}\partial_{\bar{\mu}}
\chi_{\left(2|s|+1\right)} 
\,\int_0^{L}\,dz\,\left[\,\cos{\left(|r|+|s|-1\right)kz}\,\right]
\,+\,H.C.
\label{a8}
\end{eqnarray}
where $2r+1=4l+1$, $2s+1=4p+3$ (l,p=0,1,2,....) or vice versa.
Because of the periodicity of cosine 
function
the terms in (\ref{a8}) give non-zero 
contributions after integration over $z$ only if the arguments of 
cosines are zero. This is possible only when
\begin{eqnarray}
|r|+|s|-1=0~~~\Rightarrow~~~~~r=0~~,~~s=1
~~~~\mbox{or}~~~~
s=1~~,~~r=0 \label{a10}
\end{eqnarray}
The result of $z$ integration in (\ref{a8}) is
\begin{eqnarray}
\frac{L}{2}\int \,d^4x
\left[\,i\bar{\chi}_{1}\gamma^{\bar{\mu}}\partial_{\bar{\mu}}\chi_{3}
\,+\,i\bar{\chi}_{3}\gamma^{\bar{\mu}}\partial_{\bar{\mu}}\chi_{1}
\,\right]\,+\,H.C.
\label{a11}
\end{eqnarray}
The diagonalization of (\ref{a11}) results in 
\begin{eqnarray}
&&\frac{1}{2}L
\int \,d^4x
\left[\,i\bar{\psi}\gamma^{\bar{\mu}}\partial_{\bar{\mu}}\psi
\,-\,i\bar{\tilde{\psi}}\gamma^{\bar{\mu}}\partial_{\bar{\mu}}\tilde{\psi} 
\,\right]\,+\,H.C.
\label{a12} \\
&&\psi\,=\,
\frac{1}{\sqrt{2}}\left(\,\chi_1\,+\,\chi_3\,\right)
~~,~~\tilde{\psi}\,=\,
\frac{1}{\sqrt{2}}\left(\,\chi_1\,-\,\chi_3\,\right) \label{a12a}
\end{eqnarray}
Hence there are one usual fermion and one ghost fermion in the spectrum.

The $i\bar{\chi}\gamma^{4}\partial_4\chi$ 
part of ${\cal L}_f$ reduces to  
$\sin{\frac{|n|-|m|}{2}kz}$ type of terms as a result of the action of the 
derivative operator $\partial_4$ in (\ref{a2}) (see Appendix A, part 2). 
This, in turn, 
results in odd number of sine terms (in the action) that leads to zero 
after integration over $z$. The number of modes that survive after 
integration may be increased by changing the extra dimension dependent 
conformal factor and/or the dimension of the space. For example if the 
conformal factor in (\ref{a1}) is changed to $\cos^2{kz}$ then the 
condition (\ref{a10}) is changed into $r+s-3=0$. The kinetic term 
induces no mass term in this case as well because the extra dimensional 
derivatives induce odd number of sine terms in this case as well. The same 
second term in (\ref{a2}), that is, the spin connection term also induce 
no mass term because it contains odd number of sine terms as well (see 
Appendix A, part 1). A 
similar conclusion should be expected for more complicated 
conformal terms or higher dimensional spaces. In 
other words no mass is induced for Kaluza-Klein modes through the extra 
dimensional part of the kinetic term in this model, and similar results 
are expected for more complicated situations with similar conformal terms 
and symmetries. Here I have taken $\chi_n$'s be simply given by the first 
terms in (\ref{a4c}). However taking the general form does not change the 
conclusion  because vanishing of the extra dimensional kinetic term after 
integration follows directly from the extra dimensional coordinates rather 
than the extra dimensional form of the spinor.

I have introduced an extra dimensional model where only two modes 
of Kaluza-Klein tower appear at low energies. These modes correspond to 
a fermion and a ghost fermion. These 
fermions are massless provided we do not introduce a bulk mass term 
explicitly. The ghost fermion may be identified by a Lee-Wick \cite{Lee} or 
Pauli-Villars \cite{Pauli-Villars} type regularization field. These
results are quite non-standard both in the emergence of a finite number of 
Kaluza-Klein modes and the modes higher than zero mode gaining no masses 
through the extra dimensional piece of kinetic term at low energies where 
the extra dimensions become directly unobservable. In fact this is also 
the basic tool to distinguish this scheme from the usual Kaluza-Klein 
prescription. If nature behaves in the way described here then all 
Kaluza-Klein modes of a fermion will be observed at short distances 
smaller than the size of the corresponding extra dimension 
while only a finite number of these modes will be detected at larger 
scales after they are produced (even when they are stable 
or long living so that they can travel large distances before decay). 
Moreover the coupling of fermions to other 
particles would vary nonlinearly with distance at the scales smaller than 
the size of the extra dimension since the screening effect of the 
conformal factor $\cos{kz}$ changes nonlinearly at distances below the 
size of the extra dimension. This would be another characteristic of this 
type of models. In fact one may easily find different metrics of different 
form and in different dimensions with finite number of Kaluza-Klein modes 
(obtained after integration over extra dimensions) and all with zero masses.
The aim of this study is to show the possibility of obtaining finite 
number of Kaluza-Klein modes at low energies, and the possibility of 
massless Kaluza-Klein modes higher than zero mode. So a relatively 
simple model where these properties can be observed is studied here rather 
than a detailed model that is in agreement with phenomenology. I hope 
different variations of such models with more realistic spectra may be found 
in future.

\begin{acknowledgments}
This work was supported in part by Scientific and Technical Research 
Council of Turkey under grant no. 107T235.
\end{acknowledgments}

\appendix \section{Possible Contributions to Masses Due to The Spin 
Connection and the Extra Dimensional Part of the Kinetic Term} 
\subsection{Contribution Due to Spin Connection} The vielbeins, $e^a_B$, 
corresponding to the metric, $g_{BC}$
in (\ref{a1}) 
, and those corresponding to its 
inverse $g^{BC}$ are
determined from 
\begin{equation} 
g_{BC}\,=\,\eta_{ab}\,e^a_B\,e^b_C ~~~,~~~
g^{BC}\,=\,\eta^{ab}\,e^B_a\,e^C_b ~~~,~~~
\label{apb1} \end{equation} 
where the lower indices $a$, $b$ stand for the 
tangent space of the original space (e.g. the one defined by (\ref{a1})). 
The vielbeins  
corresponding to the metric in (\ref{a1}) are found to be 
\begin{equation} 
e^a_B\,=\,\sqrt{\cos{kz}}\,\delta^a_B~~~~
e^B_a\,=\,\frac{1}{\sqrt{\cos{kz}}}\,\delta^B_a~~~~
\label{apb2} 
\end{equation} 
where $\delta^a_B$, $\delta^B_a$ are the Kronecker delta, and 
$\left(\eta_{AB}\right)$=$\mbox{diag}(1,-1,-1,-1,-1)$. In curved spaces 
the derivative term $\partial_B$ when acting on spinors is replaced by 
$D_B$ \cite{Weinberg} 
\begin{eqnarray} 
D_B&=&\partial_B\,+\,\frac{i}{2}\,J_{bc}\omega_B^{bc} \label{apb3} \\ 
&&\mbox{where}~~~J_{bc}\,=\,-\frac{i}{4}\left[\gamma_b,\gamma_c\right] 
\nonumber 
\end{eqnarray} 
Here $\gamma_{b(c)}$ are the (flat) tangent 
space Dirac gamma matrices that are related to the gamma matrices of the 
original space $\Gamma_B$ by 
\begin{eqnarray} 
\Gamma_B\,=\,e^a_B\,\gamma_a&&~~~,~~~~ 
\left\{\Gamma_B,\Gamma_C\right\}\,=\,2\,g_{AB}\,=\, 
\frac{2}{\cos{kz}}\eta_{BC}~~~,~~~ 
\left\{\gamma_a,\gamma_b\right\}\,=\,2\,\eta_{ab} \nonumber \\ 
&&B,C=0,1,2,3,4~~~,~~~~a,b=\bar{0},\bar{1},\bar{2},\bar{3},\bar{4} 
\nonumber 
\end{eqnarray}
where the bars over the integers are used to emphasize that they belong 
to the tangent space, and $\omega_B^{bc}$'s are the spin connections, that 
are given by 
\begin{equation} \omega_B^{bc}\,=\,\left[e^b_K 
\left(\frac{\partial\,e^c_P}{\partial\,x^B}\right) 
\,-\,\Gamma^F_{PB}e^b_K\,e^c_F\right]\,g^{KP} \label{apb4} \end{equation} 
where 
$\Gamma^F_{PB}$=$\frac{1}{2}g^{FG}\left(g_{PG,B}+g_{BG,P}-g_{PB,G}\right)$ 
denotes Christoffel symbols, and the commas denote the usual derivative 
with respect to that coordinate. The non-vanishing $\Gamma^F_{PB}$'s in 
the space defined by Eq.(\ref{a1}) are 
\begin{equation} 
\Gamma^\mu_{\nu4}\,=\,-\frac{k}{2}\delta^\mu_\nu
\,\tan{kz}
~,~~ 
\Gamma^4_{\mu\nu}\,=\,\frac{k}{2}\eta_{\mu\nu}\,
\tan{kz} 
~,~~ 
\Gamma^4_{44}\,=\,\frac{k}{2}\tan{kz} \label{apb5} \end{equation} 
So the 
spin connection that gives a non-zero contribution is found to be 
\begin{equation} 
\omega_\mu^{bc}\,=\,\frac{k}{2}\tan{kz}\left[\delta^b_4\delta^c_\mu\,-
\, \delta^b_\mu\delta^c_4\right] \label{apb6} 
\end{equation} 
The $\omega_4^{bc}$ element of spin connection is found to be zero.
Then 
\begin{eqnarray} 
D_\mu&=&\partial_\mu\,+\,\frac{i}{2}\,J_{bc}\omega_\mu^{bc}\,=\, 
\partial_\mu\,+\frac{k}{8}\tan{kz}\, 
\left[\gamma_{\bar{4}},\gamma_{\bar{\mu}}\right]~~~,~~~D_4\,=\,\partial_4 
\label{apb7} \\ 
\Gamma^B\,D_B&= &e^\mu_a\gamma^a\,D_\mu\,+\,e^4_a\gamma^a\,\partial_4 
\,=\,\frac{1}{\sqrt{\cos{kz}}}\,\left[\,\gamma^{\bar{\mu}}\,D_{\bar{\mu}} 
\,+\,\gamma^{\bar{4}}\,\partial_{\bar{4}}\,\right] \label{apb8} 
\end{eqnarray} 
where the upper case indices B,C,F etc. denote the 
space-time coordinates while the lower case indices a,b,c etc. and the 
indices with a bar over e.g. $\bar{B}$, $\bar{4}$ etc. denote the tangent 
space. Although there is a bar over 4 in (\ref{apb7}) that bar is omitted 
in (\ref{a2}) to simplify the notation. 
Therefore 
the result may be written a more compact form as in Eq.(\ref{a2}) where 
$\omega_4^{ab}$ gives null contribution.

After using Eqs. (\ref{apb7}) and (\ref{apb8} one obtains Eq.(\ref{a2}). 
It is evident from (\ref{a2}) 
and Eq.(\ref{apb6})
that the integration of the spin connection 
term $e^{\mu}_a\gamma^a\omega_\mu^{bc}J_{bc}$ over the extra dimension $z$ 
is proportional 
to \begin{equation} 
\int_0^{2\pi} \left(\cos{kz}\right)^2\tan{kz}\,d(kz)\;=\;0 \label{apb9} 
\end{equation} In other words the spin connection term dose not contribute 
to the masses of $\psi$, $\tilde{\psi}$ of Eq.(\ref{a12a}) at (relatively 
low energies) where the extra dimension can not be seen.

\subsection{Contribution Due to the Extra Dimensional Part of the Kinetic 
Term}

The extra dimensional part of the kinetic term for the action of the 
field $\chi$ is
\begin{eqnarray}
&&\int\,\left(\cos{kz}\right)^{\frac{5}{2}}
i\bar{\chi}\gamma^4\partial_4\chi\,d^4x\,dz\,=\,
\sum_{r,s=0}^\infty \int \,d^4x
\,i\bar{\chi}_{\left(2|r|+1\right)}\gamma^4
\chi_{\left(2|s|+1\right)}\, 
\int\,dz\,{\left(\cos{kz}\right)}^2 \nonumber \\
&&\times\,
\{\,{\left(\cos{\frac{2|r|+1}{2}kz}\,+\,\sin{\frac{2|r|+1}{2}kz}\right)}
\partial_4{\left(\cos{\frac{2|s|+1}{2}kz}\,-\,\sin{\frac{2|s|+1}{2}kz}\right)} 
\nonumber \\
&+&{\left(\cos{\frac{2|r|+1}{2}kz}\,-\,\sin{\frac{2|r|+1}{2}kz}\right)}
\partial_4{\left(\cos{\frac{2|s|+1}{2}kz}\,+\,\sin{\frac{2|s|+1}{2}kz}\right)}
\,\}\nonumber \\
&=&
-k\sum_{r,s=0}^\infty \left(2|s|+1\right)\int \,d^4x
\,i\bar{\chi}_{\left(2|r|+1\right)}\gamma^4
\chi_{\left(2|s|+1\right)}\, 
\int\,dz\,{\left(\cos{kz}\right)}^2 \nonumber \\
&&\left[
\cos{\frac{2|r|+1}{2}kz}\,
\sin{\frac{2|s|+1}{2}kz}
\,+\,\sin{\frac{2|r|+1}{2}kz}\,
\cos{\frac{2|s|+1}{2}kz}\right]\,=\,0 \label{apb10}
\end{eqnarray}
where the $H.C.$ symbol (in Eq.(\ref{a8})) for the addition the 
Hermitian conjugate of 
(\ref{apa4}) to itself is suppressed.
So the extra dimensional piece of the kinetic term in this paper
does not contribute to the masses of $\psi$ or $\tilde{\psi}$ 
at length scales larger than the size of the extra dimension.

\section{Derivation of Eq.(\ref{a8})}
It is observed that 
\begin{eqnarray}
\mbox{as}~~~~
k\,z\,\rightarrow\,\pi\,+\,k\,z &&\nonumber \\
i)~~\mbox{if}~~~n=4l+1~&\Rightarrow&~~
\left(\cos{\frac{n}{2}kz}+\sin{\frac{n}{2}kz}\right)~~\rightarrow
\left(\cos{\frac{n}{2}kz}-\sin{\frac{n}{2}kz}\right) \nonumber \\
&&
\left(\cos{\frac{n}{2}kz}-\sin{\frac{n}{2}kz}\right)~~\rightarrow
-\left(\cos{\frac{n}{2}kz}+\sin{\frac{n}{2}kz}\right) \nonumber \\
ii)~~\mbox{if}~~~n=4l+3~&\Rightarrow&~~
\left(\cos{\frac{n}{2}kz}+\sin{\frac{n}{2}kz}\right)~~\rightarrow
-\left(\cos{\frac{n}{2}kz}-\sin{\frac{n}{2}kz}\right) \nonumber \\
&&\left(\cos{\frac{n}{2}kz}-\sin{\frac{n}{2}kz}\right)~~\rightarrow
\left(\cos{\frac{n}{2}kz}+\sin{\frac{n}{2}kz}\right) \nonumber \\
&&l=0,1,2,.....~~~~~~~~
\label{apa1}
\end{eqnarray}
The requirement that the action (\ref{a2}) be invariant under 
(\ref{a6a})requires $n=4l+1$ type of modes couple to $m=4p+3$ type of 
modes. In the light of this observation the combination that is 
invariant under (\ref{a6}) is 
\begin{eqnarray}
&&\{\,\left(\cos{\frac{2|r|+1}{2}kz}\,+\,\sin{\frac{2|r|+1}{2}kz}\right)
\left(\cos{\frac{2|r|+1}{2}kz}\,-\,\sin{\frac{2|r|+1}{2}kz}\right)
\nonumber \\
&+&\left(\cos{\frac{2|r|+1}{2}kz}\,-\,\sin{\frac{2|r|+1}{2}kz}\right)
\left(\cos{\frac{2|s|+1}{2}kz}\,+\,\sin{\frac{2|s|+1}{2}kz}\right)\} 
\nonumber \\
&&=\,2\left[
\cos{\frac{2|r|+1}{2}kz}
\,\cos{\frac{2|s|+1}{2}kz}
\,-\,\sin{\frac{2|r|+1}{2}kz}
\,\sin{\frac{2|s|+1}{2}kz} \right]
\label{apa2}
\end{eqnarray}
where
\begin{eqnarray} 
&&2|r|+1=4l+1~~\mbox{and}~~2|s|+1=4p+3 \nonumber \\
&&\mbox{or}~~~2|r|+1=4l+3~~\mbox{and}~~2|s|+1=4p+1 \nonumber \\
&&l,p=0,1,2,3,........~~~~~~~~~\label{apa3}
\end{eqnarray}
So the 4-dimensional part of $S_f$ becomes
\begin{eqnarray}
&&\sum_{r,s=0}^\infty \int \,d^4x
\,i\bar{\chi}_{\left(2|r|+1\right)}\gamma^{\bar{\mu}}\partial_{\bar{\mu}}
\chi_{\left(2|s|+1\right)}\, 
\int\,dz\,{\left(\cos{kz}\right)}^2 \nonumber \\
&&\times\,
\{\,{\left(\cos{\frac{2|r|+1}{2}kz}\,+\,\sin{\frac{2|r|+1}{2}kz}\right)}
{\left(\cos{\frac{2|r|+1}{2}kz}\,-\,\sin{\frac{2|r|+1}{2}kz}\right)} 
\nonumber \\
&+&{\left(\cos{\frac{2|r|+1}{2}kz}\,-\,\sin{\frac{2|r|+1}{2}kz}\right)}
{\left(\cos{\frac{2|s|+1}{2}kz}\,+\,\sin{\frac{2|s|+1}{2}kz}\right)}
\,\} \label{apa4}
\end{eqnarray}
where the $H.C.$ symbol (as in Eq.(\ref{apb10})) 
is suppressed.
(\ref{apa4}) after use of (\ref{apa2}) results in (\ref{a8}). It is 
evident from (\ref{a8}) that the resulting Lagrangian has the form 
required by (\ref{a5}) as well.


\bibliographystyle{plain}

\end{document}